\documentclass[journal,table,xcdraw]{IEEEtran}
\IEEEoverridecommandlockouts
\pdfoutput=1
\usepackage{cite}
\usepackage{amsmath,amssymb,amsfonts}
\usepackage{graphicx}
\usepackage{textcomp}
\usepackage{cite}
\usepackage{hyperref}
\usepackage{multirow}
\usepackage{arydshln}
\usepackage{graphicx}
\usepackage{svg}
\usepackage{subcaption}
\usepackage{cleveref}
\usepackage{algorithm}
\usepackage{algpseudocode}
\usepackage{gensymb}
\usepackage{amsmath}
\usepackage{nomencl}
\allowdisplaybreaks
\makenomenclature
\usepackage{booktabs,tabulary}
\usepackage{xurl}

\DeclareMathOperator{\EX}{\mathbb{E}}
\hypersetup{ 
	colorlinks=true,
	linkcolor=black,
	filecolor=black,      
	urlcolor=black,
	citecolor=black,
}

\def\BibTeX{{\rm B\kern-.05em{\sc i\kern-.025em b}\kern-.08em
    T\kern-.1667em\lower.7ex\hbox{E}\kern-.125emX}}

\begin{document}

\title{\textcolor{black}{Wildfire Risk-Informed Preventive-Corrective Decision Making under Renewable Uncertainty}}



\author{\IEEEauthorblockN{Satyaprajna Sahoo, \textit{Student Member, IEEE},
and 
Anamitra Pal, \textit{Senior Member, IEEE}}

\thanks{This work was supported in part by  the National Science Foundation (NSF) grant under Award ECCS-2132904.

Satya Sahoo and Anamitra Pal are with
the School of Electrical, Computer, and Energy Engineering (ECEE) at Arizona State University (ASU). Emails: sssahoo2@asu.edu; Anamitra.Pal@asu.edu

}

\vspace{-2em}
}


\maketitle

\begin{abstract}
The increasing frequency and intensity of wildfires poses severe threats to the secure and stable operation of power grids, particularly one that is interspersed with renewable generation.
\textcolor{black}{Unlike conventional contingencies, wildfires affect multiple assets, leading to  cascading outages and rapid degradation of system operability and stability.}
At the same time, the usual precursors of large wildfires, namely dry and windy conditions, are known with high confidence at least a day in advance.
Thus, a \textit{coordinated} decision-making scheme employing both day-ahead and real-time information 
\textcolor{black}{has a significant potential to mitigate}
dynamic wildfire risks in renewable-rich power systems.
Such a scheme is developed in this paper through a novel \textit{stochastic preventive-corrective cut-set and
stability-constrained unit commitment and optimal power flow formulation} that also accounts for the variability
of renewable generation.
The results obtained using a reduced 240-bus system of the US Western Interconnection
demonstrate that the proposed approach increases the resilience of power systems across multiple levels of wildfire risks while maintaining economic viability.
\end{abstract}


\IEEEpeerreviewmaketitle
\vspace{-1em}
\printnomenclature[1.3cm]


\section{Introduction}

\IEEEPARstart{W}{ildfires} 
have been steadily increasing in 
frequency and intensity over the past decade
\cite{cunningham2024increasing}. 
They particularly pose a huge risk to secure and stable
power system operation.
This is because the environmental conditions that precede most large wildfires, such as \textit{high temperature}, \textit{low humidity}, and \textit{strong winds}, also exacerbate line loading and increase likelihood of vegetation contact with power lines.
Similarly, \textit{particulate matter} from wildfires approaching 
right-of-way of lines (guided by wind) frequently result in arc-faults.
Unlike conventional faults, wildfire-related faults are harder to detect (as
they are high impedance faults), have unpredictable durations, and 
occur multiple times over a short time-period \cite{ozansoy2025temporal,sahoo2023cut}.
Managing such faults
is outside the purview of conventional contingency analysis tools
that usually
cater to a predefined
set of known asset overloads.
Further, cascading outages of lines and generators triggered by wildfire-related faults can cause power system instability and 
blackouts
\cite{savastianov2024power}.


The above-mentioned impacts of wildfires can worsen when a system has numerous renewable generation sources.
Specifically,
insufficient fault ride-through performance of inverter-based resources (IBRs) has been identified as a contributing factor to cascading outages in both the 2025 Iberian blackout and the 2019 UK blackout \cite{ENTSOE2025IncidentReport,NGESO2019LFDDReport}.
However, apart from analyzing the impact of wildfire smoke on photovoltaic output 
\cite{ali2024data},
not much work has been done on investigating how IBR-rich systems may be more susceptible to the security and stability risks posed by wildfires.
Particularly, it has been shown in \cite{jalilian2021novel} that \textit{voltage regulation} is an important problem in IBR-rich systems operating in wildfire-prone areas.

One way to mitigate the risks associated with 
wildfires and IBRs on power system operation
is by integrating security-and-stability-constraints into an optimal power flow (OPF) formulation.
However, even a 
transient stability-constrained OPF (TSCOPF) 
problem is a non-linear and non-convex $\mathrm{NP}$-hard problem \cite{yuan2021robustly}. This means that it is challenging to solve 
a security-and-stability-constrained OPF problem
in real-time under variable wildfire risks and IBR uncertainties.
At the same time, it is worth noting that the environmental precursors of large wildfires and system loading conditions, are known to a high level of confidence in the day-ahead stage \cite{greenough2025wildfire}.
Therefore, a coordinated day-ahead and real-time decision-making can be effective in managing evolving wildfire risks while maintaining secure, stable, and economic grid operation.

In accordance with this realization,
a preventive-corrective scheme for mitigating wildfire risks, including a comprehensive contingency analysis tool that ensured both (static) security against 
cascading line outages as well as (dynamic) transient stability, was proposed in \cite{sahoo2024preventive}. 
However, the uncertainty associated with wildfire risk estimation was not considered in that scheme. Moreover, 
\cite{sahoo2024preventive} 
did not account for the increased operational variability caused by 
renewable generation sources (solar, wind). 
Finally, voltage regulation in the context of IBRs was not investigated in \cite{sahoo2024preventive}.
In this paper, we build on \cite{sahoo2024preventive} by:

\begin{itemize}
    \item 
    Incorporating consideration of
    \textit{voltage regulation} of IBR buses and 
    \textit{uncertainty} associated with both IBRs as well as dynamic wildfire risks to create a more effective preventive-corrective scheme for mitigating wildfire risks.
    
    \item Proposing a novel \textit{stochastic look forward cut-set and security constrained unit commitment} (S-CSCUC) model that takes inputs from the contingency analysis tool, and proposes additional units to bring online for the future.
    
    \item Proposing a novel \textit{stochastic cut-set and stability constrained optimal power flow} (S-CSCOPF) that takes inputs from the contingency analysis tool and S-CSCUC, and determines
    real-time redispatch and load shed (if need be) for a wide range of wildfire risk scenarios.
\end{itemize}

The proposed framework is implemented on a reduced 240-bus IBR-rich Western Electricity Coordinating Council (WECC) system developed in \cite{yuan2020developing}. The results show that the proposed method is able to alleviate all wildfire-related instabilities for a diverse
range of wildfire risks. 

\section{Contingency Analysis and Uncertainty Modeling}
\label{Section2}
\begin{figure}[ht]
	\centering
	\includegraphics[width=0.485\textwidth]{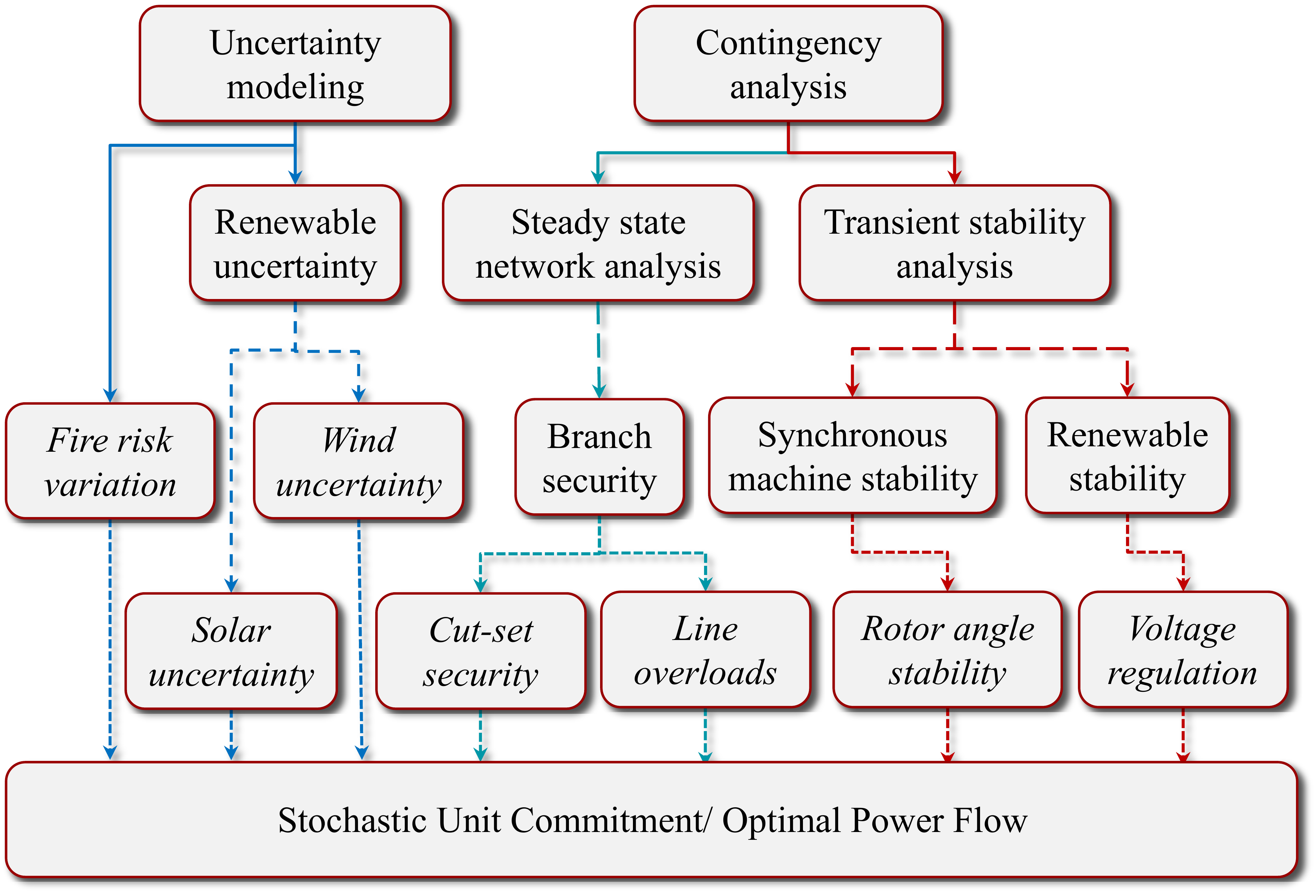}

	\caption{Contingency analysis and uncertainty modeling done in this paper}
	\label{fig:contingency_analysis_classification}
    \vspace{-2em}
\end{figure}

A distinguishing characteristic of wildfire impacts on the grid
is their potential to cause multi-asset outages that
cascade into dynamic instabilities tripping 
generating units offline \cite{nerc2018mw}. Additionally, the higher variations in IBR outputs
during wildfire events can further intensify the operational impacts of wildfires.
\textcolor{black}{Accordingly, this section analyzes key security criteria (line overloads and cascading line outages, termed \textit{cut-sets}) and stability criteria (rotor angle) 
under wildfire-induced disturbances.}
The effects of wildfires on transmission lines is alleviated by combining conventional contingency analysis with cut-set security,
while the stability of both synchronous generators as well as IBRs is assessed through their rotor angles and voltages, respectively.
Finally, uncertainty criteria for both renewable generators and wildfire risks are taken into account and converted into suitable scenarios for a stochastic UC/OPF formulation. The different criteria are shown in Fig.  \ref{fig:contingency_analysis_classification}, and 
elaborated subsequently.

\subsection{Cut-set Security}
A cut-set is 
a set of lines that divides the grid into at least two distinct regions. In certain cases, the loss of a line in a cut-set may redirect flows into
other lines making them exceed their capacity; this is called a \textit{saturated cut-set} \cite{biswas2021mitigation}.
By quickly detecting saturated cut-sets, the power system can be made secure against multiple line outages.
In this paper, a fast and scalable algorithm called \textit{feasibility test} (FT), developed in \cite{biswas2020graph}, is used to exhaustively identify all saturated cut-sets.
Once identified, they are alleviated by being incorporated into the UC/OPF model as constraints of the form:
\begin{equation}
    \sum_{K_{\mathrm{crit}}}  \phi_k \leq  T_c \quad  \forall K_{\mathrm{crit}}
\end{equation}
where $\phi_k$ is the active power flowing in line $k$ of the cut-set $K_{\mathrm{crit}}$, and $T_c$ is the transfer margin, or the aggregate amount of  power that the cut-set must shed to maintain security.

\vspace{-1em}
\subsection{Synchronous Generator Transient Stability}
For synchronous machines, transient stability is assessed in the form of \textit{rotor angle} stability, the basis of which comes from the maximum difference in the rotor angles of two consecutive generators $\delta_{\mathrm{max}}$ crossing a certain threshold:
\begin{equation}
    \mathrm{TSI} = \frac{360 - \delta_{\mathrm{max}}}{360 + \delta_{\mathrm{max}}} \times 100
    \label{TSI}
\end{equation}

In \eqref{TSI}, $\mathrm{TSI}$ is the \textit{transient stability index}, and the system is unstable if $\mathrm{TSI}\leq0$. In an unstable case, the post contingency rotor angles can be used to separate the set of synchronous machines $G_s$ into a set of critical machines $\mathrm{CM}$, and a set of non-critical machines $\mathrm{NM}$. The instability can then be corrected by shifting the appropriate amount of generation (stability margin, defined as $T_d$) from $\mathrm{CM}$ to $\mathrm{NM}$ using the \textit{integrated extended equal area criterion} (IEEAC) \cite{7395386}. This is expressed by the following constraint:
\begin{equation}
    \sum^{\mathrm{CM}}_i p_i \leq T_d
\end{equation}
where $T_d$ is computed through a linear regression (LR) model that takes pre-contingency loads $l^*$ as input and is trained in the day-ahead stage  \cite{sahoo2023cut}. Mathematically, this is written as:
\begin{equation}
    T_d = \sum_{i \in L} \theta_i l^*_i + \theta_0 = \Upsilon(l)
 \end{equation}
where $\theta$ are the weights of the LR model. This allows for faster implementation as conducting large numbers of time domain simulations (TDS) is not feasible in real-time. 

More information regarding the computation of $T_c$ and $T_d$ can be found in  \cite{sahoo2024preventive,sahoo2023cut}.

\vspace{-0.5em}
\nomenclature{\(\delta\)}{Rotor angle of synchronous machines}
\nomenclature{\(\theta\)}{Linear regressor weights}
\nomenclature{\(\Upsilon\)}{Transient stability constraint predictor model}






\vspace{-0.5em}
\subsection{Solar and Wind Output Uncertainty}
\label{sec_solar}

Solar generation is inherently uncertain due to its dependence on variable weather conditions such as solar irradiance, cloud cover, smoke, and temperature. 
To model the variations in solar generation output, a variety of distributions have been used. 
An $n$-component Gaussian mixture model (GMM) is used in this paper to model deviation in solar output ($p^r$) due to the GMM's versatility and ability to capture multimodal uncertainty characteristics \cite{shafiullah2020gaussian}:
\begin{subequations}
\begin{align}
f^s(p^r) &= \sum_{i=1}^{n} \omega_i \, \mathcal{N}\big(p^r \mid \mu_i, \sigma_i^2\big), \\
\mathcal{N}(p^r \mid \mu_i, \sigma_i^2) &= \frac{1}{\sqrt{2 \pi \sigma_i^2}} \exp\Bigg(-\frac{(p^r-\mu_i)^2}{2\sigma_i^2}\Bigg)
\end{align}
\label{eq:GMM_solar}
\end{subequations}

\vspace{-1em}
In \eqref{eq:GMM_solar}, $\mu$ and $\sigma$ are the mean and the variance vectors of the $n$ components, respectively, while $\omega$ denotes the weight vector. The expected value of the solar output is given by:  
\begin{equation}
    \mathbb{E}[p^r] = \sum_{i=1}^{n} \omega_i \, \mu_i
\end{equation}

\nomenclature{\(f^s\)}{Solar output deviation PDF}
\nomenclature{\(w\)}{Weights of the Gaussian components}
\nomenclature{\(\mu\)}{Means of the  Gaussian components}
\nomenclature{\(\sigma\)}{Variance of the Gaussian components}


\textcolor{black}{Similarly, error in wind speed estimation has been modeled using a Beta distribution \cite{wind_beta_proof}. 
Specifically, 
the stochastic variation in wind speed is captured in this paper through a Beta ($B$) distribution with parameters $o,q$:}
\begin{equation}
        f^{w}(p^r; o, q)= \frac{1}{B(o,q)} \, \epsilon^{o-1} (1-p^r)^{q-1}
        \label{eq:wind_beta}
\end{equation}
\nomenclature{\(f^w\)}{Wind output deviation PDF}

\vspace{-0.5em}

\textcolor{black}{Optimal values of the hyperparameters $(\mu, \sigma,\omega, o,q)$ are obtained empirically in Section \ref{implement} through publicly available solar irradiance and wind speed datasets.}

\vspace{-0.7em}

\subsection{\textcolor{black}{Variation in Estimation of Wildfire Risk}}
\label{sec:1a}

While extensive studies have been conducted on forecasting wildfire risk from weather, 
such analyses are often found to over-predict high wildfire risk in certain situations \cite{di2025global}. Moreover, this error increases with increase of the prediction horizon \cite{mcnorton2024global}. Hence, it is important to consider the variation in wildfire risk prediction and calculation. \textcolor{black}{Beta distributions have been used in literature to model errors in wind speed \cite{wind_beta_proof} and  atmospheric humidity \cite{price2001study}. Since these features are precedents for estimating wildfire risk,} 
the wildfire risk variation is also considered to follow a Beta distribution in this paper having parameters $o^{\prime},q^{\prime}$:
\begin{subequations}
\begin{align}
    f^{\lambda}(\lambda; o^{\prime}, q^{\prime}) &= \frac{1}{B(o^{\prime},q^{\prime})} \, \lambda^{o^{\prime}-1} (1-\lambda)^{q^{\prime}-1}, 
    && 0 < \lambda < 1, \\
    B(o^{\prime},q^{\prime}) &= \int_0^1 t^{\,o^{\prime}-1}(1-t)^{\,q^{\prime}-1} \, dt
\end{align}
\label{eq:fire_beta}
\end{subequations}

In \eqref{eq:fire_beta}, $\lambda$ is the wildfire risk, and $(o^{\prime},q^{\prime})$ are the parameters of the probability distribution function $f^\lambda$. The random variable $\lambda$ is specified to vary about its expected value ($\mathbb{E}[\lambda]$):
\begin{equation}
    \mathbb{E}[\lambda] = \frac{o^{\prime}}{o^{\prime}+q^{\prime}}
\end{equation}

\textcolor{black}{Note that the values of
$o^{\prime}, q^{\prime}$ are obtained empirically in Section \ref{implement} through publicly available wildfire risk data.}
\nomenclature{\(f^\lambda\)}{Wildfire risk probability density function (PDF)}
\nomenclature{\(o, o^{\prime}\)}{Beta distribution first parameter}
\nomenclature{\(q, q^{\prime}\)}{Beta distribution second parameter}

\section{Stochastic Preventive-Corrective Coordination with IBR Voltage Regulation} \label{Section3}

\begin{figure*}[ht]
\centering
\includegraphics[width=1\textwidth]{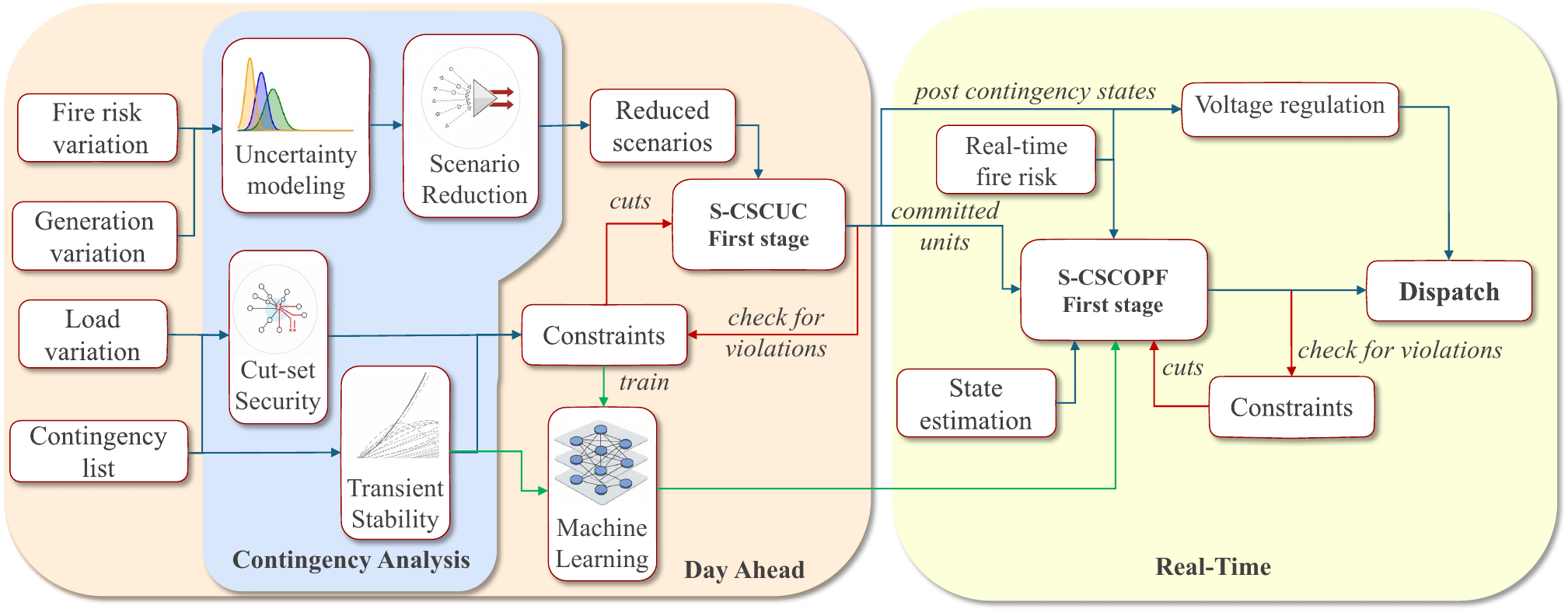}
\vspace{-1.75em}
\caption{{Implementation flowchart of the proposed risk-aware scheduling and dispatch framework  integrating
contingency analysis with day-ahead S-CSCUC and real-time S-CSCOPF formulations}}
\label{fig:implementation_flowchart}
\vspace{-1em}
\end{figure*}

\textcolor{black}{
This section employs the uncertainty modeling and contingency screening described in the previous section to create a multi-timescale stochastic preventive–corrective coordination scheme that ensures secure dispatch under dynamic wildfire conditions.
The scheme consists of a day-ahead S-CSCUC formulation and a real-time S-CSCOPF formulation as shown in Fig. \ref{fig:implementation_flowchart}, with both the formulations implemented in two stages.
The first stage of the S-CSCUC model solves the UC problem
with minimal constraints. This is followed by the second stage which is a feasibility check, where the solution is validated recursively for violations of all constraints, and subsequently added as cuts in the next iteration.
The \textit{redispatch} outputs of the UC model $p$ and load shed $l$ give insight into the post-contingency scenario in the event  of a wildfire, and is used for voltage regulation of the IBR buses. The generation commitment solution $u$ denotes the generators to bring online in the next day, and is 
an input to the S-CSCOPF model. Simultaneously, $\Upsilon$ is trained on the contingency analysis results, so that estimation of the transient stability constraint limits $T_d$ can be done quickly
in real-time (and is not dependent on multiple TDSs). 
In real-time, the state estimates and real-time wildfire risk estimates are
used with $\Upsilon$ to solve a similarly decomposed OPF formulation, with the first stage including the objective and variable limits, while 
in the second stage, the subsequent constraints, if violated, are added iteratively.}

\vspace{-1em}

\subsection{Preventive Stochastic Cut-set and Security Constrained Unit Commitment (S-CSCUC)}
\label{sec:2b}

The UC is modeled as a limited redispatch problem based on
a warm-start approach. \textcolor{black}{A redispatch formulation in place of a conventional full UC model preserves the baseline (cold-start) schedule required for the initial contingency and uncertainty analyses. Additionally, it reduces computational burden and enables integration as a plug-in layer without replacing the existing UC framework.} The objective of this UC model is to minimize the redispatch cost of the generators, considering a particular contingency or set of contingencies: 
\begin{subequations} \label{eq:main}
    \begin{align}
        \min_{{p, l, u}} \quad & h({p, l, u}) \label{eq:objective} \\
        \text{s.t.} \quad & f_1(p, p^*, u) \leq C_1 \label{eq:constraint1} \\
        & f_2(l,l^*) \leq C_2 \label{eq:constraint2} \\
        & f_3(p,l, p^*) \leq C_3 \label{eq:constraint3} \\
        & g_4(p,l) = 0 \label{eq:constraint4}\\
        & f_5(p,l) \leq T_c \label{eq:constraint5}\\
        & f_6(p) \leq T_d \label{eq:constraint6}
    \end{align}
\end{subequations}

In \eqref{eq:main}, the decision variables for the generator redispatch $p$ and the load shed $l$ indicate the \textit{deviation} from the conventional UC solution or the previous dispatch, denoted by $p^*$ and the load forecast 
denoted by $l^*$.
\textcolor{black}{Similarly, the maximum and minimum deviation $p^{\max}, p^{\min}$ is obtained from the generation capacities and the base case dispatch $p^*$.}
The objective $h(p,l,u)$ is divided into the generator redispatch cost, the load shed cost, and the startup cost. Constraints \eqref{eq:constraint1} and \eqref{eq:constraint2} bind the generator and load limits using the dispatch set-points $p^*$ and forecast $l^*$.
The branch flow limits and $N-1$ line contingency constraints are modeled in \eqref{eq:constraint3} using the dispatch line flows. The energy balance constraints are modeled in \eqref{eq:constraint4}. Constraints \eqref{eq:constraint5} and \eqref{eq:constraint6} are the cut-set and transient stability constraints obtained from contingency analyses. Eqs. \eqref{eq:constraint1} through \eqref{eq:constraint6} are linear convex formulations of the decision variables.

\nomenclature{\(p\)}{Deviation in active power output of generators}
\nomenclature{\(l\)}{Load shed}
\nomenclature{\(u\)}{Unit commitment binary variable}
\nomenclature{\(T_c\)}{Cut-set security constraint limits}
\nomenclature{\(T_d\)}{Transient stability constraint limits}

The overall goal of the UC problem is to determine the additional generators to dispatch given the uncertainty of generation and contingency realization. In real-time, these reserve units may be 
dispatched to increase the readiness level of the system.
Since the objective of the UC problem is to decide which generators to bring online to reduce operational cost, 
the problem is formulated as a robust optimization problem that looks at the worst-case post-contingency.
The 
decision variables $p$ and $l$ give an insight into what the post-contingency situation would reflect. With high penetration of renewable generation, the worst-case scenario involves consideration of the uncertainty in renewable generation output. This is captured by modifying the objective into a two-stage \textit{distributionally robust} formulation:

\vspace{-1em}
\begin{equation}
\min_{p, l, u} \; \max_{p^r} \; h(p,l,u,p^r) \quad \forall {\xi \in \Xi}\label{eq:UC_ref_obj}
\end{equation}
where $p^r$ denotes the renewable generator output subject to an uncertainty probability distribution $\xi$ in the ambiguity set $\Xi$. Theoretically, \eqref{eq:UC_ref_obj} minimizes the operational cost while trying to maximize the redispatch costs incurred by the reduction in renewable generation during a post-contingency scenario. For implementation purposes, \eqref{eq:UC_ref_obj} is reformulated into an equivalent single-stage problem, as shown below:
\begin{subequations} 
    \begin{align}
        \min_{{p, l, u}} \quad & h(p, l, u, p^r) \label{eq:UC:obj:small:1}\\
        \text{s.t.} \quad & f_1(p, p^*, u) \leq C_1  \label{eq:UC:small:1}\\
        & f_2(l,l^*) \leq C_2  \label{eq:UC:small:2}\\
        & f_3(p,l, p^*, p^r) \leq C_3  \quad \forall \; \xi \in \Xi \label{eq:UC:small:3}\\
        & g_4(p,l, p^r) = 0 \label{eq:UC:small:5}\\
        & f_5(p,l, p^r) \leq T_c \quad \forall \; \xi \in \Xi \label{eq:UC:small:4}\\
        & f_6(p) \leq T_d \label{eq:UC:small:6}
    \end{align}
\end{subequations}

The objective is to minimize
the redispatch costs, while the constraints are expected to be satisfied for all uncertainty scenarios. 
The complete formulation is fleshed out as follows: 
\begin{equation}
    \begin{aligned}
    \min_{ p_{i}, l_j, u_i}  \quad &  \sum_{\forall i \in G_s} (c_i ( p_{i})^2 + d_i p_{i}) + \sum_{\forall j \in L} (m_j l_j)\\
     \quad & + \sum_{\forall i \in G_s - G_d} u_i a_i \\ 
     \text{s.t.} \quad & 
    \end{aligned}
\end{equation}
\begin{equation}
    u_i . (p_{i}^{\mathrm{min}}) \geq  p_{i} \geq u_i . (p_{i}^{\mathrm{max}}) \;  \forall i \in G_s
    \label{eq:UC:big:1}
\end{equation}
\begin{equation}
    u_i = 1 \quad \quad \quad \forall i \in G_d
    \label{eq:UC:big:1.2}
\end{equation}
\begin{equation}
    \label{eq:UC:big:1.3}
    l_j^{\mathrm{min}} \geq  l_j \geq l_j^{\mathrm{max}} \quad \forall j \in L
\end{equation}
\begin{equation}
    \label{eq:UC:big:3}
\begin{aligned}
     \quad & \phi_e^{\mathrm{min}} \leq \sum_{\forall i \in G_s} \mathrm{PTDF}^r_{e,i}  p_{i} + \sum_{\forall j \in G_r} \mathrm{PTDF}^r_{j,i}  p^r_{i} |_\xi \\ 
     \quad & - \sum_{\forall j \in L} \mathrm{PTDF}^r_{e,j} l_j \leq \phi_e^{\mathrm{max}}  \quad \forall e \in S_B \quad \forall \xi \in \Xi
\end{aligned}
\end{equation}
\begin{equation}
\label{eq:UC:big:3.2}
\begin{aligned}
    \quad & \sum_{\forall i \in G_s} (\mathrm{PTDF}^r_{e,i} + \mathrm{LODF}_{e,k} \mathrm{PTDF}^r_{k,i})  p_{i}\\
    + \quad & \sum_{\forall i \in G_r} (\mathrm{PTDF}^r_{e,i} + \mathrm{LODF}_{e,k} \mathrm{PTDF}^r_{k,i})  p^r_{i} |_\xi \\
     - \quad & \sum_{\forall j \in L} (\mathrm{PTDF}^r_{e,j} + \mathrm{LODF}_{e,k} \mathrm{PTDF}^r_{k,j}) l_j\\
    \leq \quad & \phi_e^{\mathrm{max}} - \phi_e^* + (\mathrm{LODF}_{e,k} \phi_k^*) \quad \forall e \in S_B \quad \forall \xi \in \Xi
\end{aligned}
\end{equation}
\begin{equation}
\label{eq:UC:big:3.3}
\begin{aligned}
    \quad & \sum_{\forall i \in G_s} (\mathrm{PTDF}^r_{e,i} + \mathrm{LODF}_{e,k} \mathrm{PTDF}^r_{k,i})  p_{i}\\
    + \quad & \sum_{\forall i \in G_r} (\mathrm{PTDF}^r_{e,i} + \mathrm{LODF}_{e,k} \mathrm{PTDF}^r_{k,i})  p^r_{i} |_\xi \\
     - \quad & \sum_{\forall j \in L} (\mathrm{PTDF}^r_{e,j} + \mathrm{LODF}_{e,k} \mathrm{PTDF}^r_{k,j}) l_j\\
    \geq \quad & \phi_e^{\mathrm{min}} - \phi_e^* + (\mathrm{LODF}_{e,k} \phi_k^*) \quad \forall e \in S_B \quad \forall \xi \in \Xi
\end{aligned}
\end{equation}
\begin{equation}
    \label{eq:UC:big:5}
    \sum_{\forall i \in G_s}  p_{i} +\EX \left[ \sum_{\forall i \in G_r}  p^r_{i} \right] = \sum_{\forall j \in L}  l_j
\end{equation}
\begin{equation}
\label{eq:UC:big:4}
    \begin{aligned}
        \quad & \sum_{\forall i \in G_s} (\sum_{\forall u \in K_{\mathrm{crit}}} \mathrm{PTDF}_{u,i}) p_{i} \\
        + \quad & \sum_{\forall i \in G_r} (\sum_{\forall u \in K_{\mathrm{crit}}} \mathrm{PTDF}_{u,i})  p^r_{i}|_\xi \\
        - \quad & \sum_{\forall j \in L} (\sum_{\forall u \in K_{\mathrm{crit}}} \mathrm{PTDF}_{u,j})  l_j\\
        \leq \quad &  T_{c} ;  \quad \forall K_{\mathrm{crit}} \in \kappa_{\mathrm{crit}} \; \forall \xi \in \Xi
    \end{aligned}
\end{equation}
\begin{equation}
\label{eq:UC:big:6}
    \sum_{\forall i \in \mathrm{CM}} p_{i} \leq T_{d}
\end{equation}

\nomenclature{\(p^r\)}{Deviation in renewable generation output}
\nomenclature{\(\xi\)}{Uncertainty scenario}
\nomenclature{\(\Xi\)}{Uncertainty scenario set}
\nomenclature{\(G_s\)}{Set of synchronous machines}
\nomenclature{\(c\)}{Quadratic cost coefficients }
\nomenclature{\(d\)}{Composite cost coefficients }
\nomenclature{\(a\)}{No load cost coefficients }
\nomenclature{\(b\)}{Linear cost coefficients}
\nomenclature{\(L\)}{Set of all loads}
\nomenclature{\(m\)}{Load shed coefficients}
\nomenclature{\(G_d\)}{Set of dispatched generators}
\nomenclature{\(\phi\)}{Branch active power flow}
\nomenclature{\(S_B\)}{Set of all branches}
\nomenclature{\(K_{\mathrm{crit}}\)}{Identified saturated cut-set}
\nomenclature{\(\kappa_{\mathrm{crit}}\)}{Set of all identified saturated cut-sets}

In the above-mentioned equations, PTDF and LODF are the power transfer distribution factor and the load outage distribution factor, respectively. Further,
the composite cost coefficient $d_i$ is computed from the quadratic and linear cost coefficients as $d_i = b_i + 2c_ip^*_i$. The redispatch limit constraints \eqref{eq:UC:small:1}-\eqref{eq:UC:small:2} are denoted through \eqref{eq:UC:big:1}-\eqref{eq:UC:big:1.3}. The line limit and $N-1$ security constraints \eqref{eq:UC:small:3} are expressed in \eqref{eq:UC:big:3}-\eqref{eq:UC:big:3.3}, in which the line flows $\phi_i$ is obtained
from 
$p^*$.
The energy balance constraint \eqref{eq:UC:small:5} is expanded in \eqref{eq:UC:big:5}. Finally, the cut-set constraints \eqref{eq:UC:small:4} and transient stability constraints \eqref{eq:UC:small:6} are expressed
in \eqref{eq:UC:big:4} and \eqref{eq:UC:big:6}, respectively.


\subsection{Corrective Stochastic Cut-set and Security Constrained Optimal Power Flow (S-CSCOPF)}

The generators committed in the day-ahead stage are dispatched based on an OPF formulation.
The key difference here is that the OPF formulation not only accounts for the renewable generation uncertainty, but also for the wildfire risks.
To consider these impacts, 
a \textit{chance-constrained} approach is developed as shown below:
\begin{subequations}
\begin{align}
\min_{p, l} \quad &  h(p,l,p^r)  \quad \\
s.t \quad & \mathbb{P}_{\xi \in \Xi} \big( f_3(p,l, p^r, p^*) \leq C_3 \big) \geq  \epsilon  \\
\quad & \mathbb{P}_{\xi \in \Xi} \big( f_5(p,l, p^r) \leq \lambda T_c \big) \geq  \epsilon \\
\quad & \mathbb{P}_{\xi \in \Xi} \big( f_6(p) \leq \lambda T_d \big) \geq  \epsilon 
\end{align}
\end{subequations}
where $\lambda$ is a measure of the  real-time wildfire risk.
The scenario set $\Xi$ captures a wide range of extreme operating conditions, including wildfire-induced events that may require redispatch actions. A confidence level is specified such that the cut-set and transient stability constraints are satisfied with a probability of at least $\epsilon$, i.e., the constraints remain satisfied in no less than $(100\times\epsilon)$\% of all scenarios.
By modifying $\lambda$, the system readiness to a wildfire situation can be increased/decreased. 
For implementation, the OPF problem is reformulated as:
\begin{subequations} 
    \begin{align}
        \min_{{p, l}} \quad & h(p, l,p^r) \label{eq:OPF:small:obj}\\
        \text{s.t.} \quad & f_1(p, p^*) \leq C_1 \label{eq:OPF:small:1}\\
        & f_2(l,l^*) \leq C_2  \label{eq:OPF:small:2} \\
        & f_3(p,l, p^*, p^r) - C_3 \leq  u^{\prime} M \quad \forall \; \xi \in \Xi \label{eq:OPF:small:3}\\
        & g_4(p,l, p^r) = 0 \label{eq:OPF:small:5} \\
        & f_5(p,l, p^r) - \lambda T_c  \leq u^{\prime} M  \quad \forall \xi \in \Xi \label{eq:OPF:small:4}\\
        & f_6(p) - \lambda T_d  \leq u^{\prime} M \quad \forall \xi \in \Xi \label{eq:OPF:small:6}\\
        & \frac{1}{N_\xi}\sum_\xi u^{\prime} \leq 1 - \epsilon
    \end{align}
\end{subequations}
where $N_\xi$ is size of the scenario set. 
The cut-set and transient stability chance constraints are reformulated via a sample average approximation (SAA) method \cite{kim2014guide}, using binary variables $u^{\prime}$ and an arbitrarily large constant $M$. This ensures that in the required confidence interval, the constraints are satisfied, and in the other scenarios, the constraints are relaxed.
The complete formulation
is shown below:
\begin{equation}
    \begin{aligned}
    \min_{ p, l}  \quad &  \sum_{\forall i \in G_s} (c_i ( p_{i})^2 + d_i p_{i}) + \sum_{\forall j \in L} (m_j l_{j}) \\
     \text{s.t.} \quad & 
    \end{aligned}
\end{equation}
\begin{equation}
    p_{i}^{\mathrm{min}} \geq  p_{i} \geq p_{i}^{\mathrm{max}} \;  \forall i \in G_s 
    \label{eq:OPF:big:1}
\end{equation}
\begin{equation}
    \label{eq:OPF:big:1.3}
    l_j^{\mathrm{min}} \geq  l_{j} \geq l_j^{\mathrm{max}} \quad \forall j \in L 
\end{equation}
\begin{equation}
    \label{eq:OPF:big:5}
    \sum_{\forall i \in G_s}  p_{i} +   \sum_{\forall i \in G_r}  p^r_{i} = \sum_{\forall j \in L}  l_{j} \quad \forall \xi in \Xi
\end{equation}
\begin{equation}
    \label{eq:OPF:big:3.1}
\begin{aligned}
     \quad & \sum_{\forall i \in G_s} \mathrm{PTDF}^r_{e,i}  p_{i} + \sum_{\forall j \in G_r} \mathrm{PTDF}^r_{j,i}  p^r_{i, \xi} \\ 
     \quad & - \sum_{\forall j \in L} \mathrm{PTDF}^r_{e,j} l_{j}  - \phi_e^{\mathrm{max}} \leq  u^{\prime}_\xi M \\ \quad & \forall e \in S_B \quad \forall \xi \in \Xi
\end{aligned}
\end{equation}
\begin{equation}
    \label{eq:OPF:big:3.1.2}
\begin{aligned}
     \quad &  \sum_{\forall i \in G_s} \mathrm{PTDF}^r_{e,i}  p_{i} + \sum_{\forall j \in G_r} \mathrm{PTDF}^r_{j,i}  p^r_{i, \xi} \\ 
     \quad & - \sum_{\forall j \in L} \mathrm{PTDF}^r_{e,j} l_{j} - \phi_e^{\mathrm{min}} \geq -  u^{\prime}_\xi M \\
     \quad & \forall e \in S_B  \quad \forall \xi \in \Xi
\end{aligned}
\end{equation}
\begin{equation}
\label{eq:OPF:big:3.2}
\begin{aligned}
    \quad & \sum_{\forall i \in G_s} (\mathrm{PTDF}^r_{e,i} + \mathrm{LODF}_{e,n} \mathrm{PTDF}^r_{n,i})  p_{i}\\
    + \quad & \sum_{\forall i \in G_r} (\mathrm{PTDF}^r_{e,i} + \mathrm{LODF}_{e,n} \mathrm{PTDF}^r_{n,i})  p^r_{i, \xi} \\
     - \quad & \sum_{\forall j \in L} (\mathrm{PTDF}^r_{e,j} + \mathrm{LODF}_{e,n} \mathrm{PTDF}^r_{n,j}) l_{jk}\\
    - \quad & \phi_e^{\mathrm{max}} + \phi_e^* + (\mathrm{LODF}_{e,n} \phi_n^*) \leq u^{\prime}_\xi M \\ \quad & \forall e \in S_B \quad \forall \xi \in \Xi
\end{aligned}
\end{equation}
\begin{equation}
\label{eq:OPF:big:3.3}
\begin{aligned}
    \quad & \sum_{\forall i \in G_s} (\mathrm{PTDF}^r_{e,i} + \mathrm{LODF}_{e,n} \mathrm{PTDF}^r_{n,i})  p_{i}\\
    + \quad & \sum_{\forall i \in G_r} (\mathrm{PTDF}^r_{e,i} + \mathrm{LODF}_{e,n} \mathrm{PTDF}^r_{n,i})  p^r_{i, \xi} \\
     - \quad & \sum_{\forall j \in L} (\mathrm{PTDF}^r_{e,j} + \mathrm{LODF}_{e,n} \mathrm{PTDF}^r_{n,j}) l_{j}\\
    - \quad & \phi_e^{\mathrm{min}} + \phi_e^* + (\mathrm{LODF}_{e,n} \phi_n^*) \geq - u^{\prime}_\xi M\\ \quad & \forall e \in S_B \quad \forall \xi \in \Xi
\end{aligned}
\end{equation}
\begin{equation}
\label{eq:OPF:big:4}
    \begin{aligned}
        \quad & \sum_{\forall i \in G_s} (\sum_{\forall u \in K_{\mathrm{crit}}} \mathrm{PTDF}_{u,i}) p_{i} \\
        + \quad & \sum_{\forall i \in G_r} (\sum_{\forall u \in K_{\mathrm{crit}}} \mathrm{PTDF}_{u,i})  p^r_{i,\xi} \\
        - \quad & \sum_{\forall j \in L} (\sum_{\forall u \in K_{\mathrm{crit}}} \mathrm{PTDF}_{u,j})  l_{j} \\
        - \quad & \lambda_\xi T_c  \leq u^{\prime}_\xi M;  \quad \forall K_{\mathrm{crit}} \in \kappa_{\mathrm{crit}} \; \forall \xi \in \Xi
    \end{aligned}
\end{equation}
\begin{equation}
\label{eq:OPF:big:6} 
    \sum_{\forall i \in \mathrm{CM}} p_{i} -  \lambda_\xi T_d \leq u^{\prime}_\xi M \quad \forall \xi \in \Xi
\end{equation}
\begin{equation}
\label{eq:OPF:big:7} 
    \frac{1}{N_\xi} \sum u^{\prime}_{\xi} \leq 1 - \epsilon
\end{equation}
\nomenclature{\(\epsilon\)}{Confidence interval}
\nomenclature{\(\lambda\)}{Real-time wildfire risk}
\nomenclature{\(u^{\prime}\)}{Binary variable chance constraints}
\nomenclature{\(M\)}{Big M}
\nomenclature{\(N_\xi\)}{Number of scenarios in uncertainty set}


\vspace{-0.5em}

The outcome of the S-CSCOPF formulation is the generator redispatch $p$ and load shed $l$ for every scenario in the scenario set. Based on the wildfire risk in real-time, 
$\lambda$ can be varied 
to obtain 
a solution that the power utility is comfortable with.

\vspace{-0.5em}

\subsection{IBR Voltage Regulation}

During active wildfire risk scenarios,
the primary concern w.r.t IBRs is maintaining the
bus voltages \textcolor{black}{of the IBR subsystem $S_{\mathrm{IBR}}$} 
within acceptable limits \cite{jalilian2021novel}. The permissible voltage ride-through (VRT) limits are defined by IEEE Std. 2800-2022 \cite{IEEE2800-2022}, which 
specify the range and duration of voltage deviations that IBRs must withstand without tripping during grid disturbances.
\textcolor{black}{
The decision variables of the S-CSCUC model $(p, l)$ are
used to assess bus voltages through TDSs.
In the day-ahead stage, these voltages offer an insight into potential voltage violations if a wildfire does manifest.
In real-time, the wildfire risk forecasts are used
to determine regulation using an iterative sensitivity-based method, as outlined in Algorithm \ref{alg:voltage_regulation}. 
The method
identifies buses which have voltage violations by running a TDS for the post-contingency S-CSCUC solution for each case $\xi$ in the contingency set $\Xi$.
After obtaining the voltage sensitivity matrix $\gamma$, the voltage violations are rectified by utilizing available set of voltage control devices $\mathcal{J}$ in the network; the availability is determined using a breadth-first search (BFS)-based approach. 
Success verification is performed by running another TDS post-correction. 
}

\begin{algorithm}
\color{black}
\caption{Iterative Sensitivity-Based Voltage Regulation}
\label{alg:voltage_regulation}
\begin{algorithmic}[1]
\State \textbf{Inputs:} Subsystem $\mathcal{S}_{\mathrm{IBR}}$, Case $\xi$, Contingency set $\Xi$
\State \textbf{Output:} Updates for voltage control devices (shunts)
\For{each contingency $\xi \in \Xi$}
    \State Execute TDS and extract post-contingency states $V_{\mathrm{bus}}$
    \State Identify violation set 
    $\mathcal{V} = \{ i \in \mathcal{S}_{\mathrm{IBR}} \;|\; v_i(t) \in V_{bus}(t)$ \Statex \hspace{\algorithmicindent}
    violates the voltage--time ride-through envelope$\}$
    \If{$\mathcal{V}$ is not empty}
        \For{each violating bus $i \in \mathcal{V}$}
            \State Identify available shunts $\mathcal{J}$ 
            using BFS 
            \State Compute sensitivities $\gamma_{ij} = \frac{\partial V_i}{\partial Q_j}$ via Jacobian 
            \State Rank shunts $j \in \mathcal{J}$ by magnitude $|\gamma_{ij}|$ 
            \While{$v_i$ in violation $\&$ steps in $\mathcal{J} \neq \emptyset$}
                \State Select $j = \arg\max_{j \in \mathcal{J}} |\gamma_{ij}|$
                \State Susceptance change $\Delta \zeta_j = \frac{v_{\mathrm{target}} - v_i}{v_i \cdot \gamma_{ij}}$
                \State Map $\Delta \zeta_j$ to nearest available discrete step 
                \If{solved and $v_i$ is within limits}
                    \State Update $v_i$ and re-evaluate $\mathcal{J}$
                \Else
                    \State Revert shunt $j$ and remove from $\mathcal{J}$
                \EndIf
            \EndWhile
        \EndFor
    \EndIf
\EndFor
\end{algorithmic}
\end{algorithm}

\nomenclature{\(v_i\)}{ Voltage at bus $i$}
\nomenclature{\(S_{\mathrm{IBR}}\)}{Set of all IBR buses}
\nomenclature{\(\mathcal{V}\)}{Set of all violations}
\nomenclature{\(\mathcal{J}\)}{Set of shunts}
\nomenclature{\(\gamma\)}{Reduced Jacobian sensitivities}
\nomenclature{\(\zeta_i\)}{Susceptance of shunt $i$}

\section{Implementation}
\label{implement}
\begin{figure}[ht]
    \centering
    \begin{tabular}{cc}
        \begin{subfigure}[b]{0.22\textwidth}
            \centering
            \includegraphics[width=\linewidth]{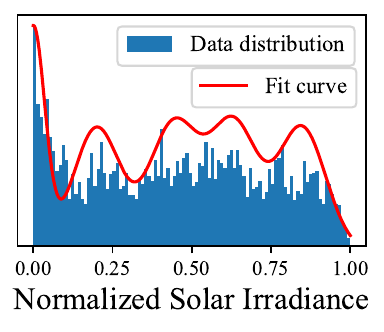}
            \vspace{-2em}
            \caption{Solar Irradiance Variation}
            \label{fig:a}
        \end{subfigure} &
        \begin{subfigure}[b]{0.22\textwidth}
            \centering
            \includegraphics[width=\linewidth]{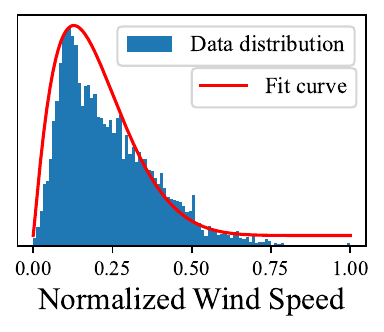}
            \vspace{-2em}
            \caption{Wind Speed variation}
            \label{fig:b}
        \end{subfigure} \\

        \begin{subfigure}[b]{0.22\textwidth}
            \centering
            \includegraphics[width=\linewidth]{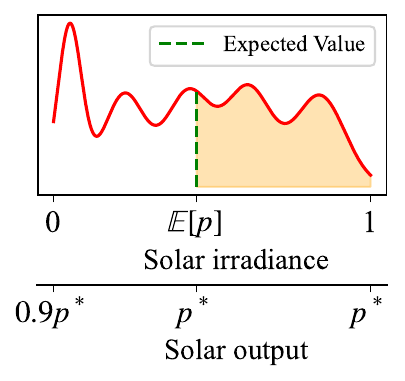}
            \vspace{-2em}
            \caption{Solar Output Variation}
            \label{fig:c}
        \end{subfigure} &
        \begin{subfigure}[b]{0.22\textwidth}
            \centering
            \includegraphics[width=\linewidth]{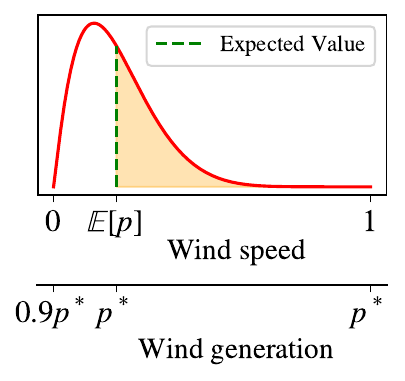}
            \vspace{-2em}
            \caption{Wind generation variation}
            \label{fig:d}
        \end{subfigure}
    \end{tabular}
    \vspace{-0.5em}
    \caption{Uncertainty analysis for solar and wind generation. Figs. \ref{fig:a} and \ref{fig:b} show the irradiance and wind speed variations, and Figs. \ref{fig:c} and \ref{fig:d} map these distributions to the solar and wind generation output}
    \label{fig:four-uncertainty_renewables}
    \vspace{-1.25em}
\end{figure}

The proposed contingency analysis and the S-CSCUC/OPF models are implemented on a reduced 240-bus WECC
system, originally developed for analyzing dynamic behavior of IBR-rich grids \cite{yuan2020developing}.
The system has a resource mix that can simulate up to 78\% renewable penetration, and the geographical embedding helps create more interpretable contingency scenarios.
The system
has a total installed solar and wind generation capacity of 9.9 GW and 21 GW, respectively, and their corresponding solar irradiance and wind speed data are sourced from NREL’s National Solar Radiation Database (NSRDB) [31]. The distribution parameters for data mapping—namely, $\sigma$, $\mu$, $o$, $q$, and $\omega$—are obtained by using an appropriate curve-fitting technique that satisfies \eqref{eq:GMM_solar} and \eqref{eq:wind_beta}, respectively. The resulting
derived stochastic models are used to generate ensembles of irradiance and wind speed deviations, which are then employed to produce ensembles of generation deviations (shown in Fig. \ref{fig:four-uncertainty_renewables}).
\textcolor{black}{These ensembles of deviation data are then independently sampled to generate a 
dataset of $\approx10,000$ samples for each instance in the network. This is followed by scenario reduction using k-nearest neighbors (kNN) 
to create 500 representative scenarios.}


Empirical studies 
show that solar output varies about 1\% during clear skies, and up to 10\% on average during cloudy days \cite{Habte2021PVVariability}
.
Similarly, wind output generally varies between 2-10\% \cite{Liu2023WindPower} for most regions and for typical weather conditions. Hence, the uncertainty for solar and wind generation output is capped at 10\% below their dispatch values.

\begin{figure}[t]
    \vspace{-0.25em}
    \centering
    \begin{tabular}{cc}
        \begin{subfigure}[b]{0.22\textwidth}
            \centering
            \includegraphics[width=\linewidth]{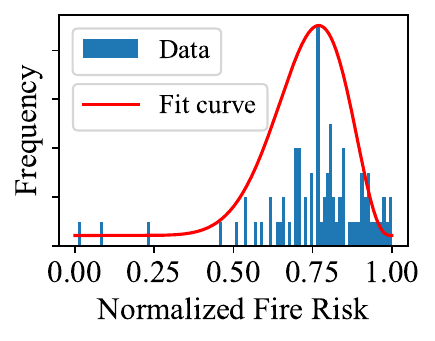}
            \vspace{-2em}
            \caption{Fire risk variation}
            \label{fig:fire_risk_variations}
        \end{subfigure} &
        \begin{subfigure}[b]{0.22\textwidth}
            \centering
            \includegraphics[width=\linewidth]{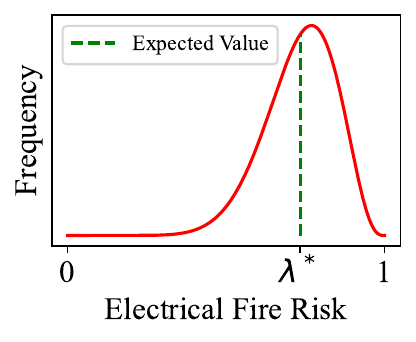}
            \vspace{-2em}
            \caption{Distribution example}
            \label{fig:fire_risk_regions}
        \end{subfigure}
    \end{tabular}
    \vspace{-0.5em}
    \caption{Uncertainty analysis for wildfire risk. Fig. \ref{fig:fire_risk_variations} shows the variation data for the risk of a particular line. Fig. \ref{fig:fire_risk_regions} maps the fit distribution to the electrical fire risk $\lambda$}
    \label{fig:fire_risk_figures}
\end{figure}

Wildfire risk data in the California region of the WECC system was mapped on the power lines of that region using \cite{rhodes2023wildfire},
and this
data is used for uncertainty analysis. Fig. \ref{fig:fire_risk_figures} shows the Beta distribution obtained when the model was fit to the risk data of the lines in the contingency set (details in Table \ref{table:contingency_details}). After obtaining the parameters $o^{\prime}, p^{\prime}$ from the data, the distribution is mapped to the electrical fire risk using a piecewise linear function,
where the expected value of the distribution is mapped onto the real-time estimated wildfire risk $\lambda^*$. The regions from 0 to  
the expected value is  mapped linearly to the region between 0 and $\lambda^*$, while the region from the expected value to 1 is linearly mapped to $\lambda^*$ to 1.


The contingency analysis including the transient stability studies were performed in $\mathrm{PSSE^\circledR}$, while the optimization models were formulated and solved using $\mathrm{Gurobi}$ and $\mathrm{Pandapower}$. 
All computational analyses done in this study were performed using \texttt{Python} on a workstation equipped with an Intel(R) Xeon(R) Gold 6246R CPU operating at 3.40 GHz with 16 physical cores and 32 logical processors, 32 GB of system memory, and an NVIDIA Quadro RTX 5000 GPU.

\vspace{-0.5em}

\section{Results}\label{Section4}

\subsection{Contingency Analysis}

\begin{figure}[hb]
    \vspace{-0.25em}
    \centering
    \begin{tabular}{cc}
        \begin{subfigure}[b]{0.22\textwidth}
            \centering
              \includegraphics[width=\textwidth]{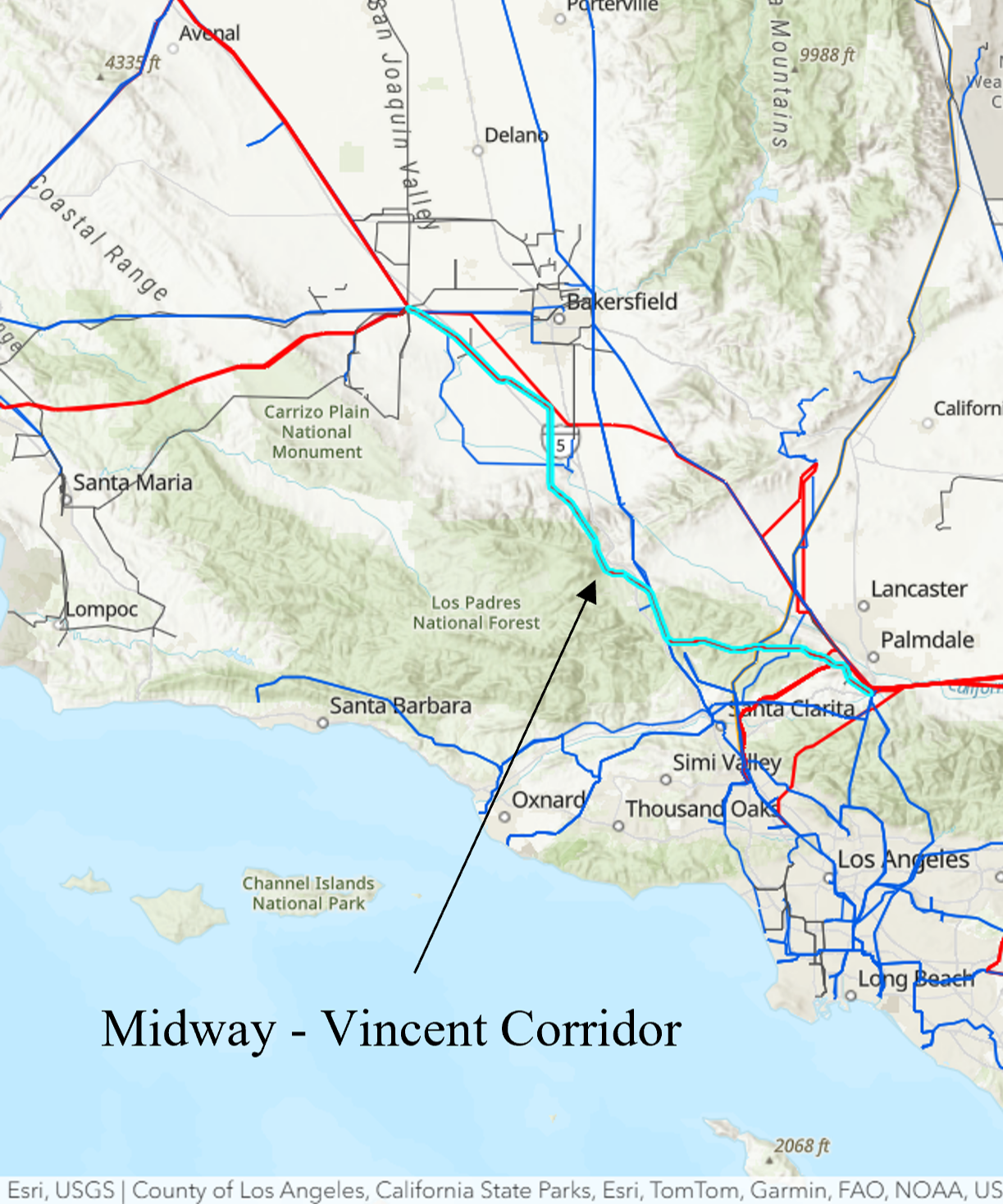}
              \caption{Location of the contingency list \cite{esri_map}}
              \label{fig:lines_location}
        \end{subfigure} &
        \begin{subfigure}[b]{0.22\textwidth}
          \centering
          \includegraphics[width=\textwidth]{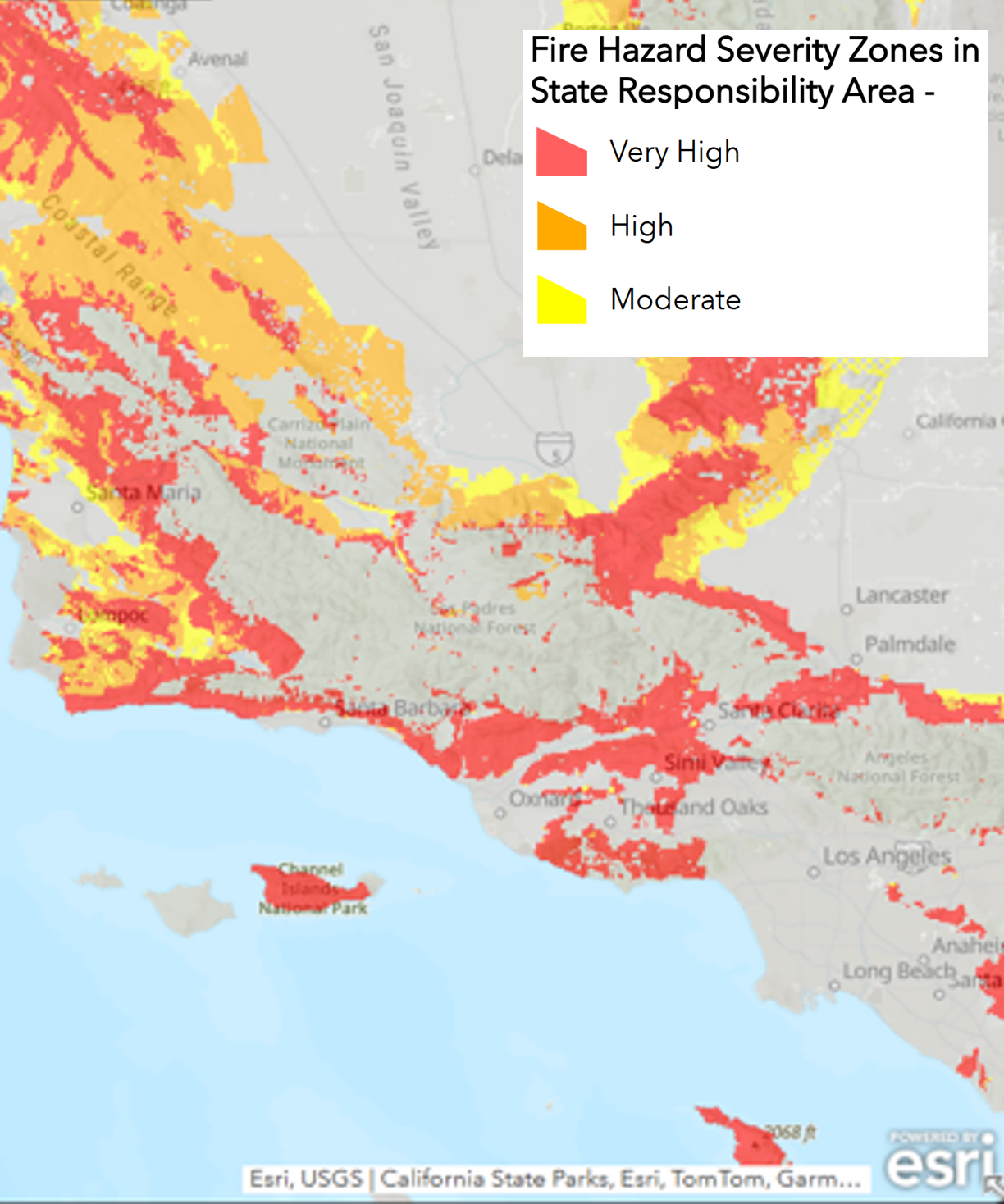}
          \caption{Fire risk in the same area \cite{firehazard2023}}
          \label{fig:cali_fire_risk}
        \end{subfigure}
    \end{tabular}
    \vspace{-0.5em}
    \caption{Geo-referencing the contingency list location. Wildfire risk in the same region (in April 2024) shows high risk in the region the lines run through.}
    \label{fig:Midway_vincent corridor}
\end{figure}

The Midway-Vincent transmission corridor (Path 26) located in the 240-bus WECC system is chosen for contingency analysis.
This corridor consists of three lines that connect northern California to southern California.
Cross-referencing the location and wildfire risk in the area shows that 
the corridor is located in a wildfire-prone area
(see Fig. \ref{fig:Midway_vincent corridor}), making it a realistic case study for the analysis conducted in this paper.
Multiple arc-faults followed by a sequential outage of the lines in the Midway-Vincent corridor is used to simulate the impacts of the wildfire.
Details of the transmission corridor as well as the contingency impacts are given in Table \ref{table:contingency_details}. 

\begin{table}[ht]
\vspace{-0.25em}
\caption{Contingency Impacts}
\vspace{-0.75em}
\centering
\begin{tabular}{|l|p{3.195cm}|}
\hline
\textbf{Property}                                                             & \textbf{Value}                                 \\ \hline
Contingency corridor & {[}(3896, 3897), (3894, 3895), (3892, 3893){]} \\ \hline
\begin{tabular}[c]{@{}c@{}}Power flowing through corridor (MW)\end{tabular} & 2000.82 \\ \hline
Cut-sets identified & [(2404,3893), (3894,3895), (3896, 3897)]\\ \hline
Cut-set transfer margin (MW) & [-305.28]\\ \hline
Unstable synchronous machine & 3831 \\ \hline
TS transfer margin (MW) &  -1635.42 \\ \hline
Unstable IBRs & [4131, 4132, 6433] \\ \hline
Total unstable generation (MW) & 2555.0 \\ \hline
\end{tabular}
\label{table:contingency_details}
\vspace{-0.75em}
\end{table}

Simulating a wildfire in the region causes line and cut-set overloads, synchronous generator instability, and IBR voltage violations.
Overall, the contingency analysis results are summarized as follows: \textit{An unchecked wildfire affecting this corridor leads
to cascading line outages causing 2,000
MW of redirected power flows, and 
2.5 GW of generation outage (and eventual load shed).} To alleviate these vulnerabilities, the contingency analysis tool proposes at least 305 MW reduction in aggregate power flowing through the cut-set, and a generation shed of at least 1,635 MW in the $\mathrm{CMs}$.

Next, the IBR voltage is regulated by
using switched shunt devices as explained in Algorithm \ref{alg:voltage_regulation}.
The post-contingency solution from the S-CSCUC identifies the IBR buses with voltage violations.
Variable shunt devices closest to these buses are then switched in discrete steps: capacitors are added to raise low voltages, and reactors are added (or capacitors removed)
to lower high voltages. After each adjustment, the power flow is recomputed to evaluate the impact on the network, and the process is repeated until post-contingency IBR voltages are within the allowable limits 
The switching sequence is selected to minimize the number of operations while maintaining voltage within prescribed limits.

\begin{figure}[htbp]
    \vspace{-1em}
    \centering
    \begin{tabular}{cc}
        \begin{subfigure}[b]{0.22\textwidth}
          \centering
          \includegraphics[width=\textwidth]{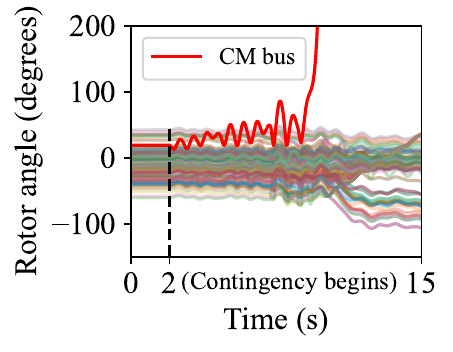}
          \vspace{-1.5em}
          \caption{TDS without control}
          \label{fig:sub1}
        \end{subfigure} &
        \begin{subfigure}[b]{0.22\textwidth}
          \centering
          \includegraphics[width=\textwidth]{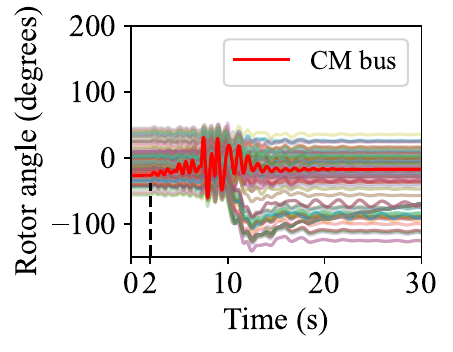}
          \vspace{-1.75em}
          \caption{TDS with control}
          \label{fig:sub2}
        \end{subfigure}
    \end{tabular}
    \vspace{-0.75em}
    \caption{Rotor angle stability for the 240-bus system}
    \label{fig:240_rotor_angles}    
    \vspace{-1.5em}
\end{figure}

\begin{figure}[htbp]
    \centering
    \begin{tabular}{cc}
        \begin{subfigure}[b]{0.22\textwidth}
            \centering
              \includegraphics[width=\textwidth]{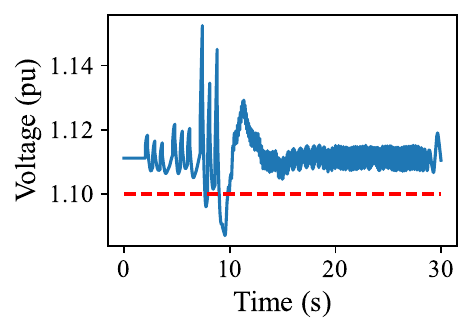}
              \vspace{-1.5em}
              \caption{Voltage of bus 6433 without regulation}
              \label{fig:ibr1}
        \end{subfigure} &
        \begin{subfigure}[b]{0.22\textwidth}
          \centering
          \includegraphics[width=\textwidth]{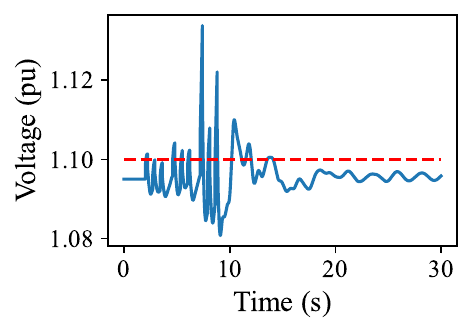}
          \vspace{-1.75em}
          \caption{Voltage of bus 6433 with regulation}
          \label{fig:ibr2}
        \end{subfigure}
    \end{tabular}
    \vspace{-0.75em}
    \caption{IBR voltage regulation for the 240-bus system}
    \label{fig:ibr_stability_images}
    \vspace{-1.5em}
\end{figure}

Performing these alleviation actions leads to stable generation and power flows, depicted in Figs. \ref{fig:240_rotor_angles} and \ref{fig:ibr_stability_images}. Fig. \ref{fig:240_rotor_angles} shows the rotor angles of the generators in the system. The red line
corresponds to the unstable generator at bus 3831. The contingencies begin at 2 seconds, and the unstable generator is seen to quickly deviate from the other generators in Fig. \ref{fig:sub1}. 
Performing the corrective action suggested in Table \ref{table:contingency_details}, the same generator can be seen to be in synchronism with the other generators post-disturbance (see Fig. \ref{fig:sub2}). 
For the IBRs, a similar behavior is observed in Fig. \ref{fig:ibr1} (for bus 6433), where the voltages violate the high VRT limits (greater than 1.0 s for voltage $>$ 1.1 p.u.) while the violation is rectified post-correction, as shown in Fig. \ref{fig:ibr2}.

\subsection{Dispatch Results}
\label{DispatchRes}

The PDFs of renewable variation and wildfire risk uncertainty are sampled randomly to generate $\approx$10,000 scenarios. 
Scenario reduction is then performed 
to obtain a smaller representative dataset $\Xi$ of
{500} distinct scenarios. The implementation results of S-CSCUC is shown in Table \ref{tab:UC_results}. Three out of six generators not dispatched are now committed. In the post-contingency scenario, two loads are shed, with the total amount of load shed being 154 MW. 

\begin{table}[ht]
\vspace{-0.25em}
\centering
\caption{S-CSCUC Results}
\vspace{-0.5em}
\begin{tabular}{|l|c|}
\hline
\textbf{Property}               & \textbf{Value} \\ \hline
Additional generators committed & 3              \\ \hline
Post-contingency redispatch cost ($\times 10^6 $ units)         & 7.13           \\ \hline
Total loads shed                & 2              \\ \hline
Total load shed (MW)            & 153.57         \\ \hline
Iterations                      & 4              \\ \hline
Time to solve (s)               & 6.38           \\ \hline
Optimality gap (\%)             & 0.0005         \\ \hline
\end{tabular}
\label{tab:UC_results}
\vspace{-0.25em}
\end{table}

In real-time, the expected value of the electrical fire risk $\lambda^*$
is varied according to the estimated wildfire risk based on environmental and other factors. The S-CSCOPF results for three different cases based on varying levels of risk is shown in Table \ref{tab:OPF_results}. In the table, $\lambda^*=0.3$ is considered to be a relatively low risk, which means that the cut-set and transient stability margins are set to 30\% of the true value for most scenarios in $\Xi$ as compared to the third case where $\lambda^* = 0.9$. Further, the confidence interval $\epsilon$ is set to 0.95, which indicates constraint violation is allowed in only a small number (5\%) of scenarios. In all three cases, the power shed in the critical lines, saturated cut-sets, and $\mathrm{CMs}$ are seen to be increasing consistently with the increase in wildfire risk. This indicates that the resilience of the system to an oncoming threat is increased at the cost of total system operations cost (denoted in units/hr), which also increases proportionately with risk. All three cases have minimal load shed of less than 1 MW. 
\textcolor{black}{In comparison, the uncoordinated CSCOPF solution developed in \cite{sahoo2023cut} would cause much higher load shed, while the \textit{deterministic} CSCUC+CSCOPF solution developed in \cite{sahoo2024preventive} would cause considerable IBR generation loss.}


\begin{table}[ht]
\centering
\caption{S-CSCOPF Results}
\vspace{-0.5em}
\begin{tabular}{|c|ccc|}
\hline
\multirow{2}{*}{\textbf{Property}}                                           & \multicolumn{3}{c|}{\textbf{Value}}                                                                       \\ \cline{2-4} 
& \multicolumn{1}{c|}{\textbf{$\lambda^*$ = 0.3}} & \multicolumn{1}{c|}{\textbf{$\lambda^*$ = 0.5}} & \textbf{$\lambda^*$ = 0.9} \\ \hline
Number of scenarios                                                          & \multicolumn{3}{c|}{500}                                                                                  \\ \hline
\begin{tabular}[c]{@{}c@{}}Contingency list \\ power shed (MW)\end{tabular}  & \multicolumn{1}{c|}{350.59}              & \multicolumn{1}{c|}{500.41}              & 1039.84             \\ \hline
Critical Machine Shed (MW)                                                   & \multicolumn{1}{c|}{496.69}              & \multicolumn{1}{c|}{822.04}              & 1472.74             \\ \hline
Cut-set desaturation (MW)                                                    & \multicolumn{1}{c|}{121.43}              & \multicolumn{1}{c|}{200.96}              & 360.04              \\ \hline
Total loads shed                                                             & \multicolumn{1}{c|}{1}                   & \multicolumn{1}{c|}{1}                   & 1                   \\ \hline
Total load shed (MW)                                                         & \multicolumn{1}{c|}{0.3}                 & \multicolumn{1}{c|}{0.65}                & 0.82                \\ \hline
\begin{tabular}[c]{@{}c@{}}Real-time generation \\ cost ($\times 10^6$units)\end{tabular} & \multicolumn{1}{c|}{0.79}                & \multicolumn{1}{c|}{1.07}                & 2.22                \\ \hline
Iterations                                                                   & \multicolumn{1}{c|}{5}                   & \multicolumn{1}{c|}{4}                   & 4                   \\ \hline
Time to solve (s)                                                            & \multicolumn{1}{c|}{15.51}               & \multicolumn{1}{c|}{13.21}               & 14.58               \\ \hline
Optimality gap (\%)                                                          & \multicolumn{3}{c|}{0}                                                                                    \\ \hline
\end{tabular}
\label{tab:OPF_results}
\end{table}

\begin{figure}[ht]
 \vspace{-1em}
	\centering
	\includegraphics[width=0.48\textwidth]{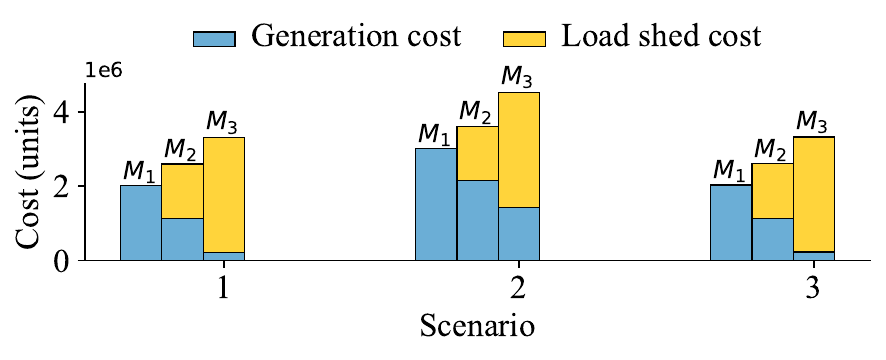}
    \vspace{-1em}
	\caption{Comparison with state-of-the-art stochastic approaches}
	\label{fig:comp_with_sota_fire}
    \vspace{-1.25em} 
\end{figure}

The comparison of how the proposed framework reduces load shed relative to alternate \textit{stochastic} approaches
is detailed in Fig. \ref{fig:comp_with_sota_fire}. 
The figure illustrates the post-contingency total operational cost when a wildfire manifests, and evaluates the performance of the proposed model ($M_1$) with two other stochastic operations models, denoted by $M_2$ and $M_3$, respectively. 
$M_2$ represents the stochastic TSCOPF (S-TSCOPF) model based on \cite{7426859},
while $M_3$ represents a stochastic economic dispatch based on \cite{wang2021data}. Comparison is done across three randomly picked 
scenarios in $\Xi$. Although $M_2$ and $M_3$ have lower generation cost than $M_1$, the load shed costs, and in turn the total costs, are much higher.
This happens because $M_2$ does not consider cut-set security and $M_3$ does not consider both cut-set security and transient stability.

\subsection{Computational Efficiency}
Total computation time is essential to ascertain the practicality and scalability of the proposed model. In order to gauge the total computational efficiency, all steps mentioned in the implementation flowchart of Fig.  \ref{fig:implementation_flowchart} must be considered. This is done as shown below.

\subsubsection{Uncertainty modeling and scenario reduction}
Model fitting and dataset generation are completed in the  order of milliseconds. Computational complexity of scenario reduction varies depending on the model utilized. In this paper, kNN is employed for scenario reduction, which identifies representative scenarios based on Euclidean distance in the feature space. The computational complexity of the kNN-based reduction is $\mathcal{O}(N^2)$, where $N$ is the original number of 
scenarios.

\subsubsection{Cut-set security analysis} 
The computational complexity of the cut-set analyses is primarily dependent on the FT algorithm, which has been proven to be efficient and scalable (see \cite{biswas2020graph,biswas2021mitigation}). For the 240-bus system, the FT is completed in the order of milliseconds.

\subsubsection{Transient stability analysis}
In the day-ahead stage, TDS runs have to be performed to determine the stability margin, and to create the dataset on which $\Upsilon$ is trained.
The run-time for one TDS for this system is of the order of magnitude of $10^0$ seconds. 
In real-time, the trained $\Upsilon$ is used which gives results within a few milliseconds.
\textcolor{black}{
\subsubsection{Voltage regulation}
IBR stability is assessed/rectified through multiple TDSs, which add a complexity of the order of $10^0$ seconds in real-time. 
}
\subsubsection{Unit commitment}
Solution time for the UC model is given in Table \ref{tab:UC_results}. The reported time of 6.38 seconds was the compute time of 4 iterations and does not include the time required for constraint evaluation and inclusion. The total computation time was of the order of $10^2$ seconds, and primarily scales with the number of binary variables. Since the redispatch model only considers the generators \textit{not} dispatched previously, the overall computation time of the model is decreased substantially.

\subsubsection{Optimal power flow}
The solution time for the OPF model given in Table \ref{tab:OPF_results} is the aggregate time over multiple iterations. The total computation time including constraint evaluation and inclusion is in the order of $10^2$ seconds. The sizes of the model and 
scenario set $\Xi$ are the primary factors influencing the overall computational burden.

\vspace{-1em}

\subsection{Sensitivity Analysis}
\begin{figure}
	\centering
	\includegraphics[width=0.44\textwidth]{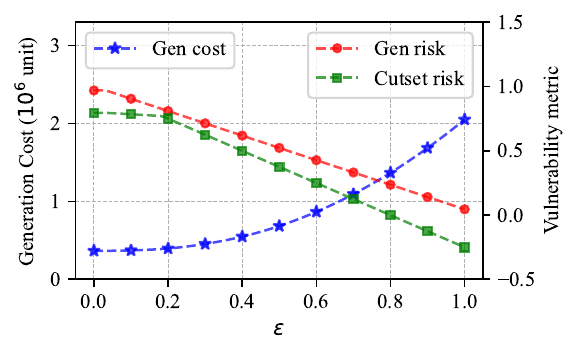}
    \vspace{-1em}
	\caption{Sensitivity: cost vs risk}
	\label{fig:epsilon_sensitivity}
    \vspace{-1em}
\end{figure}

Multiple case studies are performed to quantify the performance of the proposed method under various operating conditions.  The first case study aims to evaluate the operating cost and total system risk with the variation of $\lambda^*$. The system risk is divided into \textit{static risk} (defined as $\frac{f_5 - T_c}{T_c}$) and \textit{transient risk} (defined as $\frac{f_6 - T_d}{T_d}$). With an increase in $\lambda^*$, the static and transient risk of the system decreases, but it results in a higher system cost, as shown in Fig. \ref{fig:epsilon_sensitivity}. At $\lambda^* =1$, the wildfire is considered to have manifested, as all the security and stability margins will be set to 100\% in 
most scenarios 
in $\Xi$.

\begin{figure}
	\centering
	\includegraphics[width=0.38\textwidth]{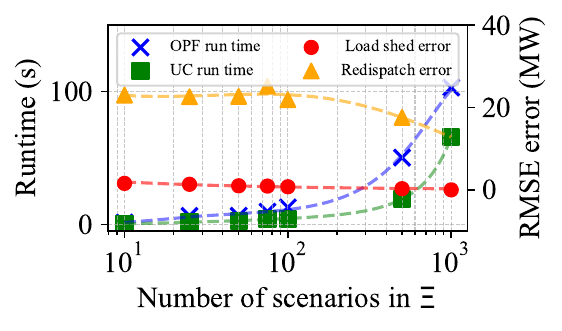}
    \vspace{-1em}
	\caption{Sensitivity: Scenario size vs. Solution time}
	\label{fig:xi_sensitivity}
    \vspace{-1em}
\end{figure}

The second case study is performed to determine the appropriate size of $\Xi$. The number of scenarios can be varied to increase or decrease the computation time, at the cost of solution accuracy. An analysis of this trade-off is illustrated in Fig. \ref{fig:xi_sensitivity}, where the solution accuracy of the UC and the OPF model is plotted against the computation time for increasing number of scenarios. From the figure, it is evident that the percentage error in generation redispatch and load shed is consistently low for most sizes of $\Xi$, while the computational complexity increases exponentially. From this analysis, it is determined that a reduced scenario set size of {500} is appropriate due to its low solve time 
and small impact on solution accuracy.

\begin{figure}[ht]
 \vspace{-2.25em}
	\centering
	\includegraphics[width=0.38\textwidth]{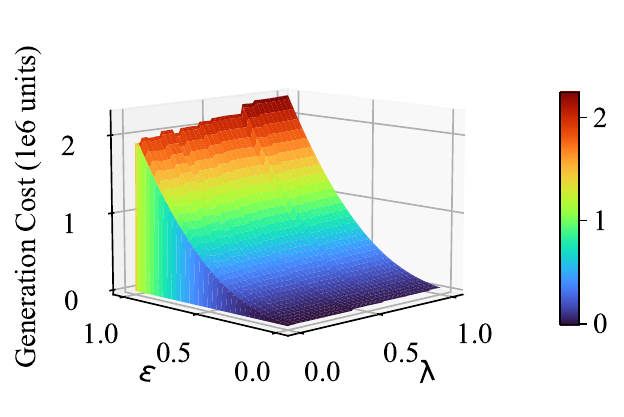}
    \vspace{-1em}
	\caption{Sensitivity analysis with $\lambda$ and $\epsilon$. $\lambda$ denotes the percentage by which the wildfire constraints should be satisfied, and $\epsilon$ denotes the percentage of scenarios for which the constraints hold.}
	\label{fig:sensitivity_analysis_eps_lam}
    \vspace{-1em}
\end{figure}

The final study involves evaluating cost
variation with total system resilience, and is essential for determining
the optimal operating point for \textit{any} risk scenario. System resilience can be quantified through $\lambda$ and the confidence interval $\epsilon$ which can be interpreted as the system tolerance to renewable uncertainty. Similarly, generation cost is also indicative of the total system risk, i.e, high operational cost is indicative of a larger amount of risk which must
be taken care of. 
Fig. \ref{fig:sensitivity_analysis_eps_lam} depicts these relations graphically. From the figure, it is observed that the \textit{operational cost exhibits higher sensitivity to renewable generation uncertainty than to wildfire risk levels} (cost increase is higher on the $\epsilon$ axis as compared to the $\lambda$ axis). This proves our earlier supposition that security or stability risks posed by wildfires can be exacerbated by renewable uncertainty and that addressing risks arising from both wildfires and renewables is necessary for ensuring grid resilience.

\section{Conclusion}



In this paper, a stochastic wildfire risk assessment and management scheme is proposed for resilient and economic operation of renewable-rich power grids under dynamic wildfire risks.
Risk assessment is performed using contingency analysis and uncertainty modeling.
The identified risks are managed from static security, transient stability, and voltage regulation perspectives through a coordinated
preventive UC and corrective OPF scheme.

The analysis conducted using
a reduced-WECC system indicates that
during periods of active wildfire risk, robust and resilient power delivery with minimal load shed is possible using the proposed scheme as compared to other 
methods. Multiple case studies demonstrate the applicability of the proposed approach
under varying conditions of active wildfire risk \textcolor{black}{and generation uncertainty}.
The results confirm the reasonable computational speed of the proposed method as well as 
identify the critical factors that impact its performance.  
The findings of this paper highlight the importance of coordinated, uncertainty-aware decision-making mechanisms for ensuring secure and resilient power system operation in the face of increasing wildfire threats. 

\vspace{-0.5em}



\bibliography{bibtex.bib}

\end{document}